\documentclass[prl,twocolumn,showpacs,floats]{revtex4}

\usepackage{graphicx}

\begin{document}

\title{Distinct Fermi Surface Topology and Nodeless Superconducting Gap in (Tl$_{0.58}$Rb$_{0.42}$)Fe$_{1.72}$Se$_2$ Superconductor}
\author{Daixiang Mou$^{1}$, Shanyu Liu$^{1}$, Xiaowen Jia$^{1}$, Junfeng He$^{1}$, Yingying Peng$^{1}$,  Lin Zhao$^{1}$, Li Yu$^{1}$,  Guodong  Liu$^{1}$, Shaolong He$^{1}$, Xiaoli Dong$^{1}$, Jun Zhang$^{1}$, Hangdong Wang$^{2}$, Chiheng Dong$^{2}$, Minghu Fang$^{2}$,  Xiaoyang Wang$^{3}$, Qinjun Peng$^{3}$, Zhimin Wang$^{3}$, Shenjin Zhang$^{3}$, Feng Yang$^{3}$, Zuyan Xu$^{3}$, Chuangtian Chen$^{3}$  and X. J. Zhou$^{1,*}$}

\affiliation{
\\$^{1}$Beijing National Laboratory for Condensed Matter Physics, Institute of Physics,
Chinese Academy of Sciences, Beijing 100190, China
\\$^{2}$Department of Physics, Zhejiang University, Hangzhou 310027, China
\\$^{3}$Technical Institute of Physics and Chemistry, Chinese Academy of Sciences, Beijing 100190, China
}
\date{January 24, 2011}
%
%

\begin{abstract}

High resolution angle-resolved photoemission measurements have been carried out to study the electronic structure and superconducting gap of the (Tl$_{0.58}$Rb$_{0.42}$)Fe$_{1.72}$Se$_2$  superconductor with a T$_c$=32 K.  The Fermi surface topology consists of  two  electron-like Fermi surface sheets around $\Gamma$ point which is distinct from that in all other iron-based compounds reported so far.   The Fermi surface around the M point shows  a nearly isotropic superconducting gap of $\sim$12 meV. The large Fermi surface near the $\Gamma$ point also shows a nearly isotropic superconducting gap of $\sim$15 meV while no superconducting gap opening is clearly observed for the inner tiny Fermi surface.   Our observed new Fermi surface topology and its associated superconducting gap will provide key insights and constraints in understanding superconductivity mechanism in the iron-based superconductors.

\end{abstract}


\maketitle

\newpage

The discovery of the Fe-based superconductors\cite{Kamihara,ZARenSm,RotterSC,MKWu11,CQJin111} has attracted much attention because it represents the second class of high temperature superconductors in addition to the copper-oxide (cuprate) superconductors\cite{Bednorz}. It is important to explore whether the  high-T$_c$ superconductivity mechanism in this new Fe-based system is conventional, parallel to that in cuprates, or along a totally new route\cite{NPReview}. Different from the cuprates where the low-energy electronic structure is dominated by Cu 3d$_{x^2-y^2}$ orbital, the electronic structure of the Fe-based compounds involves all five Fe 3d-orbitals forming multiple Fermi surface sheets: hole-like Fermi surface sheets around $\Gamma$(0,0) and electron-like ones around M($\pi$,$\pi$)\cite{DJSingh1111,Kuroki}. It has been proposed that the interband scattering between the hole-like bands near $\Gamma$ and electron-like bands near M gives rise to electron pairing and superconductivity\cite{Kuroki,FeSCMagnetic}.  An alternative picture is also proposed based on the interaction of local Fe magnetic moment\cite{J1J2Picture}.


The latest discovery of superconductivity with a T$_c$ above 30 K in a new A$_x$Fe$_{2-y}$Se$_2$ (A=K, Tl, Cs, Rb and etc.) system\cite{JGGuo,Switzerland,MHFang,GFChen} is surprising that provides new perspectives in understanding Fe-based compounds. First, it may involve Fe vacancies in the FeSe layer\cite{MHFang,GFChen,ZWang}. This is against the general belief that a perfect Fe-sublattice is essential for superconductivity in the Fe-based compounds, similar to that perfect CuO$_2$ plane is considered essential for cuprate superconductors. Second, the superconductivity of A$_x$Fe$_{2-y}$Se$_2$ is realized in  a close proximity to an antiferromagnetic semiconducting (insulating) phase\cite{MHFang,GFChen,QMSi}. This is in strong contrast to other Fe-based compounds where the parent compounds are spin-ordered metals\cite{DongSDW,RotterParent,PCDai,BFSNeutron3}. Third, the intercalation of A=(K,Tl,Cs,Rb and etc.) in A$_x$Fe$_{2-y}$Se$_2$ is expected to introduce large number of electrons into the system; this usually would lead to suppression or disappearance of superconductivity like in heavily electron-doped Ba(Fe,Co)$_2$As$_2$ system\cite{Co122}. The existence of superconductivity in A$_x$Fe$_{2-y}$Se$_2$ at such a high T$_c$ (over 30 K) with so high electron doping is unexpected. Most interestingly, band structure calculations\cite{LJZhang,IRShein,XWYan} and electronic structure measurements\cite{YZhang,TQian} all suggest that high electron doping  in A$_x$Fe$_{2-y}$Se$_2$ may lead to the disappearance of the hole-like Fermi surface sheets around $\Gamma$.  This would render it impossible for the electron scattering between the hole-like bands near $\Gamma$ and electron-like bands near M that is considered to be important for the electron pairing in Fe-based superconductors by some theoriests\cite{Kuroki,FeSCMagnetic}. Therefore, investigations of the A$_x$Fe$_{2-y}$Se$_2$ system would be important in searching for common ingredients underlying the physical properties, especially the superconductivity  of the Fe-based superconductors.

\begin{figure}[tbp]
\begin{center}
\includegraphics[width=1.0\columnwidth,angle=0]{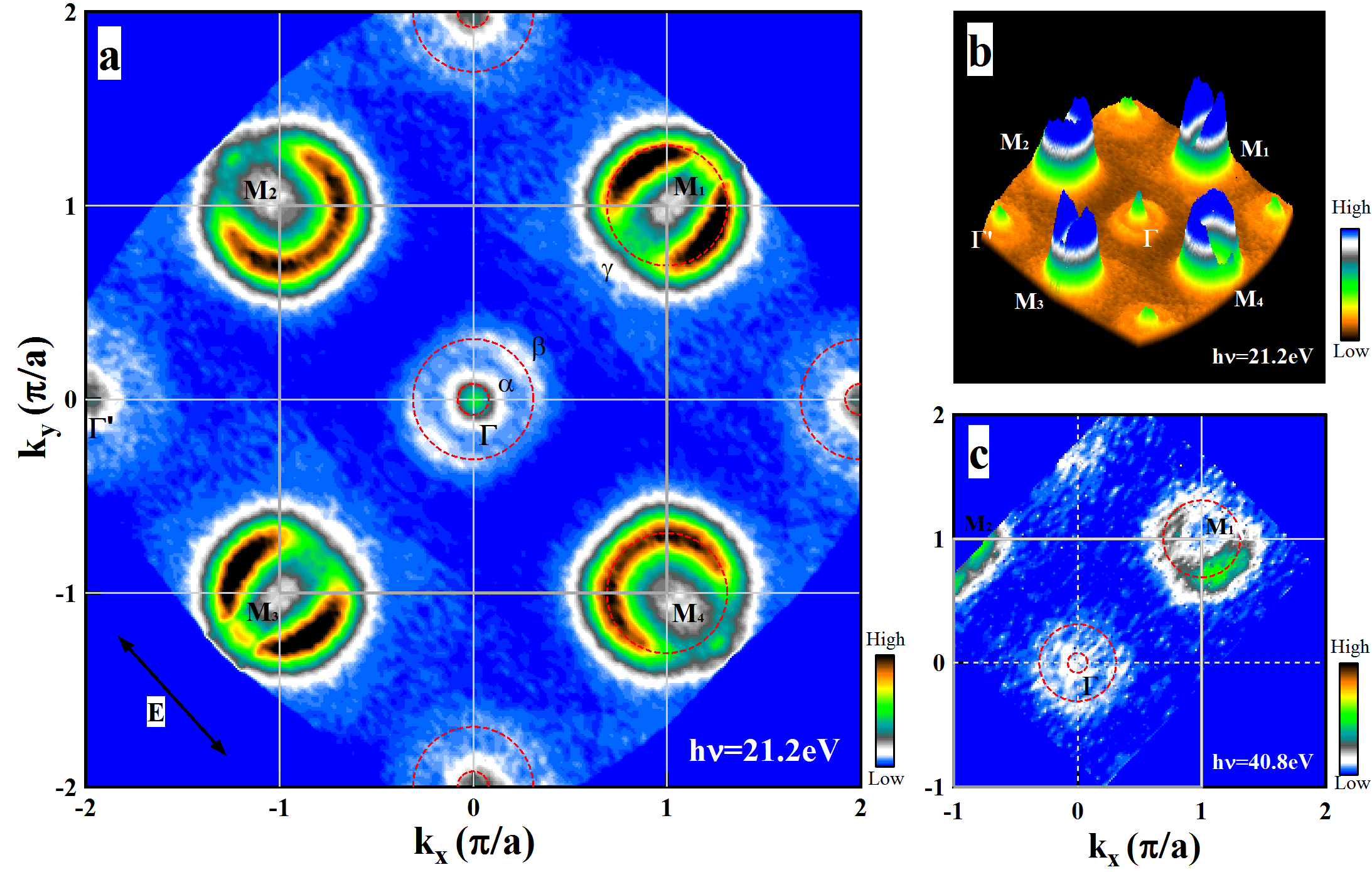}
\end{center}
\caption{Fermi surface of (Tl$_{0.58}$Rb$_{0.42}$)Fe$_{1.72}$Se$_2$ superconductor (T$_c$=32 K).  (a). Spectral weight distribution integrated within [-20meV,10meV] energy window near the Fermi level as a function of k$_x$ and k$_y$ measured using h$\nu$=21.2 eV light source.  Two Fermi surface sheets are observed around $\Gamma$ point which are marked as $\alpha$ for the inner small sheet and $\beta$ for the outer large one.  Near the M($\pi$,$\pi$) point, one Fermi surface sheet is clearly observed which is marked as $\gamma$.  (b) Three-dimensional image of Fig. 1a.  (c). Fermi surface mapping measured using h$\nu$=40.8 eV light source Although the signal is relatively weak, one can see traces of two Fermi surface sheets around $\Gamma$ and one around M.
}
\end{figure}

In this paper, we report observation of a distinct Fermi surface topology and nearly isotropic nodeless superconducting gap in (Tl$_{0.58}$Rb$_{0.42}$)Fe$_{1.72}$Se$_2$ superconductor (T$_c$=32 K) from high resolution angle-resolved photoemission (ARPES) measurements.   We have observed an electron-like Fermi surface sheet  near M($\pi$,$\pi$) and two electron-like Fermi surface sheets near $\Gamma$(0,0).  This Fermi surface topology is distinct from the hole-like Fermi surface sheets near the $\Gamma$ point found in other Fe-based compounds\cite{DJSingh1111,Kuroki} or disappearance of hole-like Fermi surface sheets near  $\Gamma$ in A$_x$Fe$_{2-y}$Se$_2$ compounds\cite{LJZhang,IRShein,XWYan,YZhang,TQian}.  We observe nearly isotropic superconducting gap around the Fermi surface sheets near $\Gamma$ ($\sim$15 meV) and M ($\sim$12 meV); no gap node is observed in both Fermi surface sheets. These rich information on this new Fe-based superconductor will provide key insights on the superconductivity mechanism in the Fe-based superconductors.


High resolution angle-resolved photoemission measurements were carried out on our lab system equipped with a Scienta R4000 electron energy analyzer\cite{GDLiu}. We use Helium discharge lamp as the light source which can provide photon energies of h$\upsilon$= 21.218 eV (Helium I) and 40.8 eV (Helium II).  The energy resolution was set at 10 meV for the Fermi surface mapping (Fig. 1a) and band structure measurements (Fig. 2) and at 4 meV for the superconducting gap measurements (Figs. 3 and 4). The angular resolution is $\sim$0.3 degree. The Fermi level is referenced by measuring on a clean polycrystalline gold that is electrically connected to the sample. The (Tl,Rb)Fe$_{2-y}$Se$_2$ crystals were grown by the Bridgeman method\cite{MHFang}. Their composition determined by using an Energy Dispersive X-ray Spectrometer (EDXS) measurement is (Tl$_{0.58}$Rb$_{0.42}$)Fe$_{1.72}$Se$_2$.  The crystals have a sharp superconducting transition at T$_{c(onset)}$=32 K with a transition width of $\sim$1 K.   The crystal was cleaved {\it in situ} and measured in vacuum with a base pressure better than 5$\times$10$^{-11}$ Torr.

\begin{figure}[b]
\begin{center}
\includegraphics[width=1.0\columnwidth,angle=0]{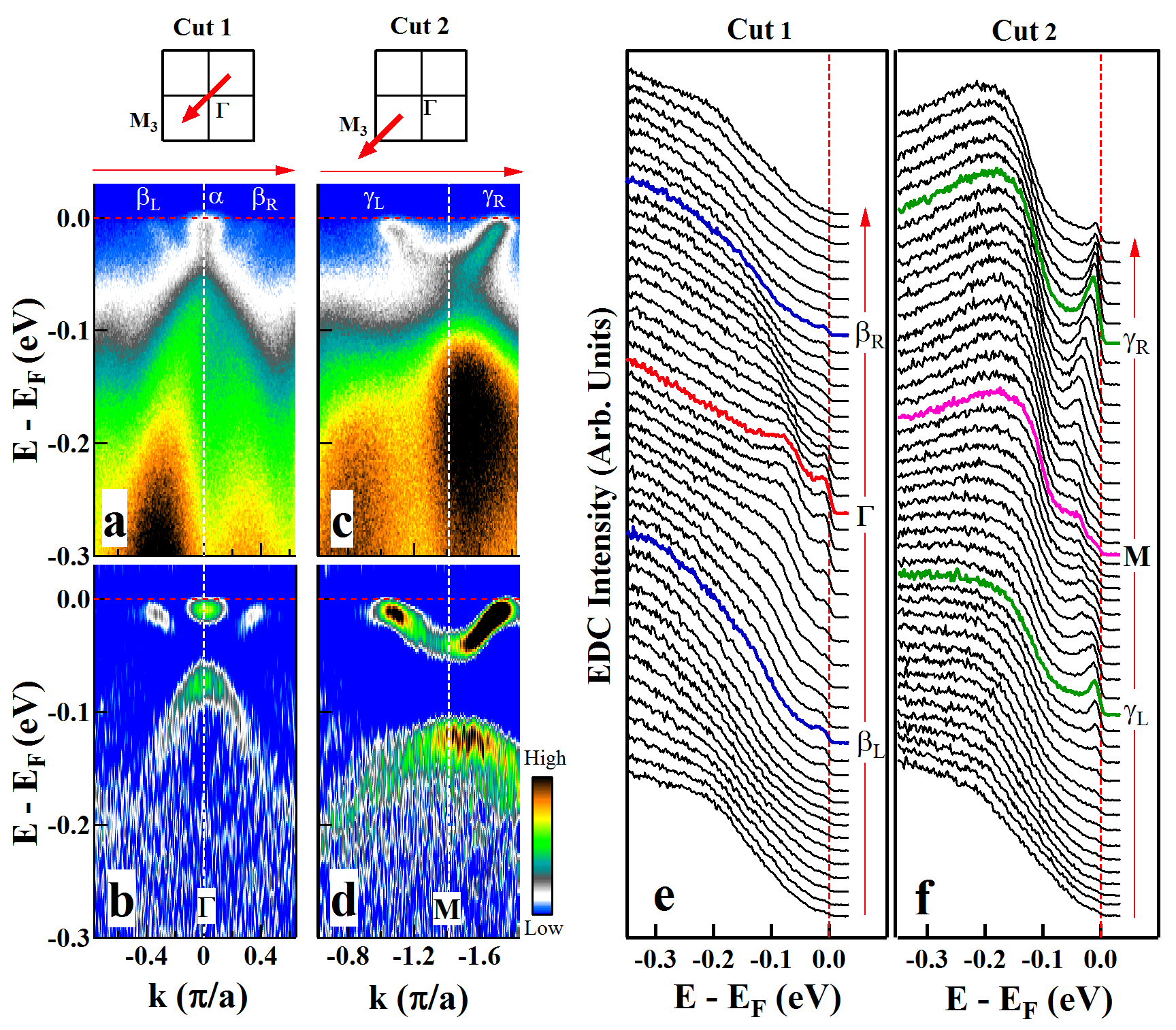}
\end{center}
\caption{Band structure and photoemission spectra of (Tl$_{0.58}$Rb$_{0.42}$)Fe$_{1.72}$Se$_2$ measured along two high symmetry cuts.
(a). Band structure along the Cut 1 crossing the $\Gamma$ point; the location of the cut is shown on top of Fig. 2a.  The $\alpha$ band and
two Fermi crossings of the $\beta$ band ($\beta_L$ and $\beta_R$) are marked.  (b). Corresponding EDC second derivative image of Fig. 2a. (c).
Band structure along the Cut 2 crossing M point; the location of the cut is shown on top of Fig. 2c. The two Fermi crossings of the $\gamma$
band ($\gamma_L$ and $\gamma_R$) are marked.  (d). Corresponding EDC second derivative image of Fig. 2c. (e). EDCs corresponding to Fig. 2a
for the Cut 1. (f). EDCs corresponding to Fig. 2c for the Cut 2.
}
\end{figure}

Fig. 1 shows Fermi surface mapping of the (Tl$_{0.58}$Rb$_{0.42}$)Fe$_{1.72}$Se$_2$ superconductor covering multiple Brillouin zones.  The band structure along two typical high symmetry cuts are shown in Fig. 2.   An electron-like Fermi surface is clearly observed around M($\pi$,$\pi$) (Fig. 1a, Figs. 2c and 2d). This Fermi surface (denoted as $\gamma$ hereafter) is nearly circular with a Fermi momentum (k$_F$) of 0.35 in a unit of $\pi$/a (lattice constant a=3.896 $\AA$).    The Fermi surface near the $\Gamma$ point consists of two sheets. The inner tiny pocket (denoted as $\alpha$) is electron-like with a band bottom barely touching the Fermi level ( Figs. 2a and 2b for the Cut 1). The outer larger Fermi surface sheet around $\Gamma$ (denoted as $\beta$) (Fig. 1a) is electron-like (Figs. 2a and 2b) with a Fermi momentum of 0.35 $\pi$/a.
 

The observation of two electron-like Fermi surface sheets, $\alpha$ and $\beta$, around $\Gamma$ in (Tl$_{0.58}$Rb$_{0.42}$)Fe$_{1.72}$Se$_2$ is distinct from that observed in other Fe-based compounds where hole-like pockets are expected around the $\Gamma$ point\cite{DJSingh1111,Kuroki}. It is also different from the band structure calculations\cite{LJZhang,IRShein,XWYan,YZhang,TQian} and previous ARPES measurements\cite{YZhang,TQian} on A$_x$Fe$_{2-y}$Se$_2$ that suggest disappearance of hole-like Fermi surface sheets near $\Gamma$ because of the lifted chemical potential due to a large amount of electron doping.

\begin{figure}[tbp]
\begin{center}
\includegraphics[width=1.0\columnwidth,angle=0]{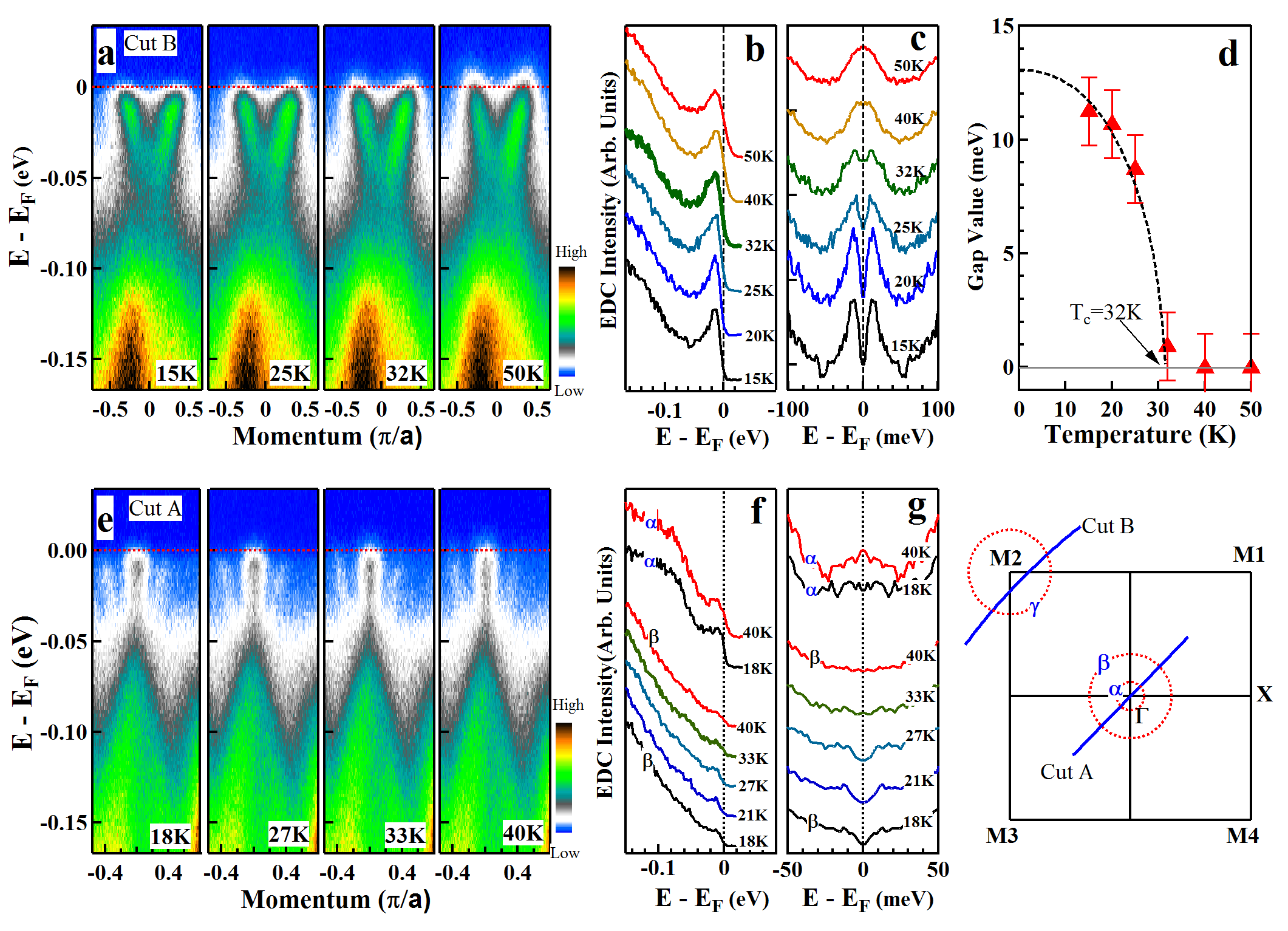}
\end{center}
\caption{Temperature dependence of energy bands and superconducting gap near $\Gamma$ and M points. (a).  Photoemission images along the Cut B (bottom-right inset). (b). Photoemission spectra at the Fermi crossing of $\gamma$ Fermi surface  and their corresponding symmetrized spectra (c) measured at different temperatures. (d). Temperature dependence of the superconducting gap. The dashed line is a BCS gap form.  (e). Photoemission images along the Cut A (bottom-right inset) at different temperatures.  The original EDCs at the $\Gamma$ point and at the Fermi crossing of $\beta$ Fermi surface measured at different temperatures are shown in (f) and their corresponding symmetrized EDCs are shown in (g).
}
\end{figure}

One immediate question is on the origin of the electron-like $\beta$ band around $\Gamma$. The first possibility is whether it could be a surface state. While surface state on some Fe-based compounds like the ``1111" system was observed before\cite{HYLiu}, it has not been observed in the ``11"-type Fe(Se,Te) system\cite{FeSTARPES}. The second possibility is whether the $\beta$ band can be caused by the folding of the electron-like $\gamma$ Fermi surface near M. It is noted that the Fermi surface size, the band dispersion, and the band width of the $\beta$ band at $\Gamma$ is similar to that of the $\gamma$ band near M.  A band folding picture would give a reasonable account for such a similarity if there exists a ($\pi$,$\pi$) modulation in the system that can be either structural or magnetic. An obvious issue with this scenario is that, in this case, one should also expect the folding of the $\alpha$ band near $\Gamma$ onto the M point; but such a folding is not observed at the M point (Fig. 1a and Fig. 2c).  The third possibility is whether the measured $\beta$ sheet is a Fermi surface at a special k$_z$ cut. Although the Fermi surface at $\Gamma$ is absent in TlFe$_2$Se$_2$ from the band structure calculations\cite{LJZhang},  there is a 3-dimensional Fermi pocket that is present near the zone center at k$_z$=$\pi$$\slash$c when x is close to 1 in K$_x$Fe$_2$Se$_2$\cite{IRShein} and Cs$_x$Fe$_2$Se$_2$\cite{XWYan}.  We note that the electron doping in (Tl$_{0.58}$Rb$_{0.42}$)Fe$_{1.72}$Se$_2$ is lower than that of (K,Cs)Fe$_2$Se$_2$. Also we observed similar $\beta$ Fermi surface at different photon energies (Fig. 1a and Fig. 1c) which corresponds to different k$_z$.  The final resolution of this possibility needs further detailed photon energy dependent measurements.


The clear identification of various Fermi surface sheets makes it possible to investigate the superconducting gap in this new superconductor.  We start first by  examining the superconducting gap near the M point.  Fig. 3a shows the photoemission images along the Cut B near M (its location shown in the bottom-right inset of Fig. 3) at different temperatures. The corresponding photoemission spectra (energy distribution curves, EDCs) on the Fermi momentum at different temperatures are shown in Fig. 3b.   To visually inspect possible gap opening and remove the effect of Fermi distribution function near the Fermi level, we have symmetrized these original EDCs to get spectra in Fig. 3c, following the procedure that is commonly used in high temperature cuprate superconductors\cite{MNorman}.   For the $\gamma$ pocket near M, there is a clear gap opening at low temperature (15 K), as indicated by an obvious dip at the Fermi energy in the symmetrized EDCs (Fig. 3c). With increasing temperature, the dip at E$_F$ is gradually filled up and is almost fully filled above T$_c$=32 K. The gap size at different temperatures is extracted from the peak position of the symmetrized EDCs or fitted by the phenomenological formula\cite{MNorman} (Fig. 3d); it is $\sim$11 meV at 15 K. The temperature dependence of the gap size roughly follows the BCS-type form (Fig. 3d).   Similar temperature dependent measurements of the superconducting gap were also carried out along the $\Gamma$-M cut near $\Gamma$ (Figs. 3e-g). The Fermi crossing on the $\beta$ Fermi surface also displays a clear superconducting gap in the superconducting state which is closed above T$_c$ (lower curves in Figs. 3f and 3g). For the peculiar tiny $\alpha$ pocket near $\Gamma$, we do not find signature of clear superconducting gap opening below T$_c$ (upper curves in Figs. 3f and 3g).


Now we come to the momentum-dependent measurements of the superconducting gap.  For this purpose we took high resolution Fermi surface mapping (energy resolution of 4 meV) of  the $\gamma$ pocket at M (Fig. 4a) and the $\beta$ pocket at $\Gamma$ (Fig. 4b).  Fig. 4c shows photoemission spectra around the $\gamma$ Fermi surface measured in the superconducting state (T= 15 K); the corresponding symmetrized photoemission spectra are shown in Fig. 4d.  The extracted superconducting gap (Fig. 4g) is nearly isotropic with a size of (12$\pm$2) meV.  The superconducting gap around the $\beta$ Fermi surface near $\Gamma$ is also nearly isotropic with a size of (15$\pm$2) meV (Figs. 4e, 4f and 4g).

\begin{figure}[tbp]
\begin{center}
\includegraphics[width=1.0\columnwidth,angle=0]{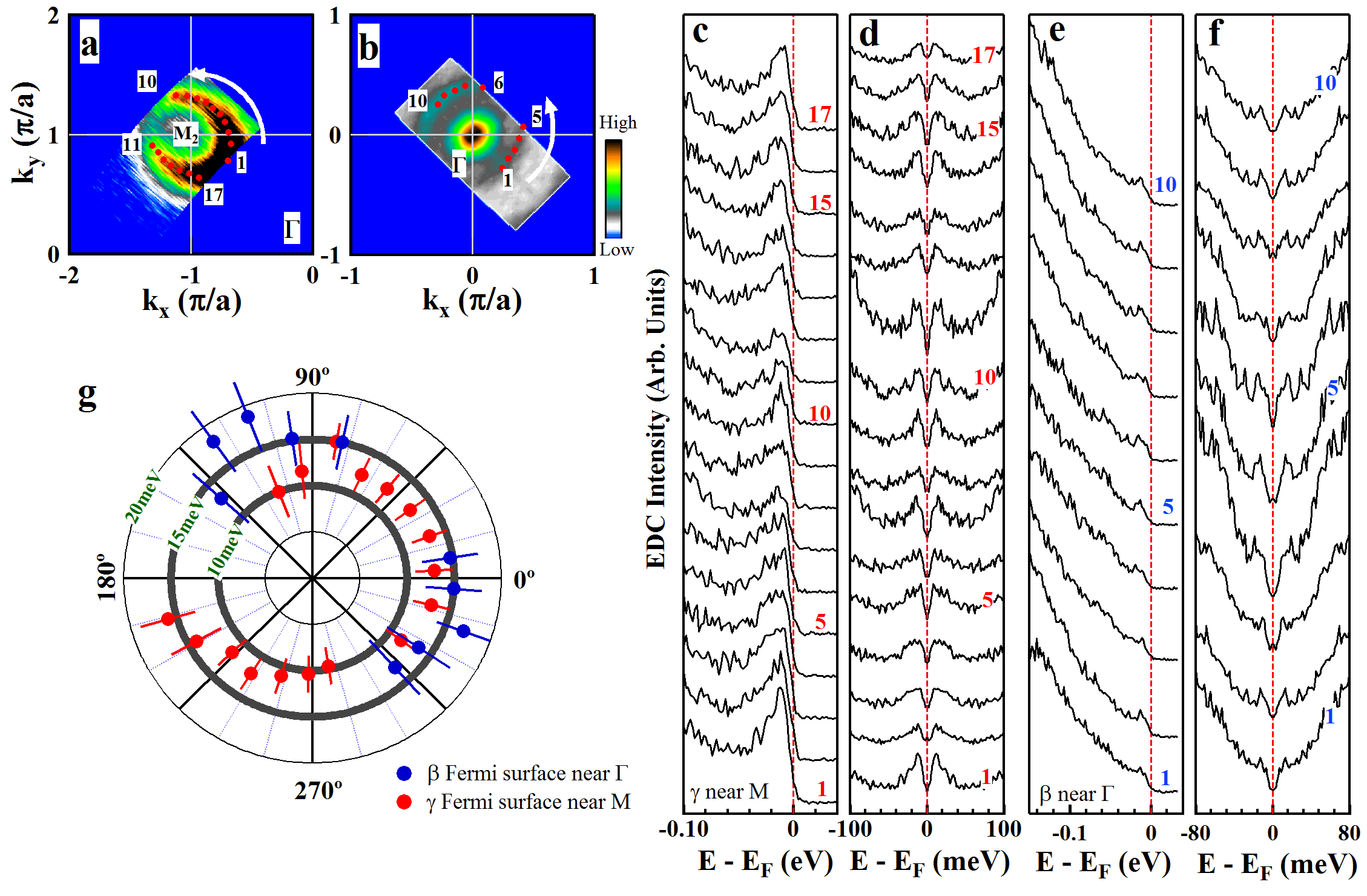}
\end{center}
\caption{Momentum dependent superconducting gap along the $\gamma$ and the $\beta$ Fermi surface sheets measured at T=15 K.  Fermi surface mapping near M (a) and near $\Gamma$ (b) and the corresponding Fermi crossings marked by red circles. (c). EDCs along the $\gamma$ Fermi surface and their corresponding symmetrized EDCs (d).  (e). EDCs along the $\beta$ Fermi surface and their corresponding symmetrized EDCs (f).   (g). Momentum dependence of the superconducting gap along the $\gamma$ Fermi surface sheet (red circles) and along the $\beta$ Fermi surface sheet (blue circles).
}
\end{figure}


The observation of a distinct Fermi surface topology in (Tl$_{0.58}$Rb$_{0.42}$)Fe$_{1.72}$Se$_2$ has important implications to the understanding of superconductivity in Fe-based superconductors.  The realization of high T$_c$ in this new superconductor with a distinct Fermi surface topology is helpful to sort out key electronic structure ingredient that is responsible for superconductivity.  With the electron-like $\beta$ Fermi surface present in (Tl$_{0.58}$Rb$_{0.42}$)Fe$_{1.72}$Se$_2$, the possibility of electron scattering between the $\Gamma$ Fermi surface sheet(s) and the  M Fermi surface sheet(s),  proposed by some theories to account for superconductivity in the Fe-based superconductor\cite{Kuroki,FeSCMagnetic}, cannot be ruled out. However, the electron scattering between two electron-like bands may have different effect on the electron pairing from that between an electron-like band and a hole-like band.  The nearly isotropic superconducting gap on the $\beta$ and $\gamma$ Fermi surface sheets, together with the absence of gap nodes, appears to favor s-wave superconducting gap symmetry in (Tl$_{0.58}$Rb$_{0.42}$)Fe$_{1.72}$Se$_2$. This is similar to that in (Ba$_{0.6}$K$_{0.4}$)Fe$_2$As$_2$\cite{122Gap} and NdFeAsO$_{0.9}$F$_{0.1}$\cite{Kaminski1111}.  The gap size of 12 (for $\gamma$) and 15 meV (for $\beta$)  gives a ratio of 2$\Delta$/kT$_c$=9 and 11, respectively,  which is significantly larger than the traditional BCS weak-coupling value of 3.52 for an s-wave gap.   This indicates that this new superconductor is at least in the strong coupling regime in terms of the BCS picture. These will put strong constraints on various proposed gap symmetries and the underlying pairing mechanisms for the iron-based superconductors.


In summary, we have identified a distinct Fermi surface topology in the new (Tl$_{0.58}$Rb$_{0.42}$)Fe$_{1.72}$Se$_2$ superconductor that is different from all other Fe-based superconductors reported so far. Near the $\Gamma$ point, two electron-like Fermi surface sheets are observed that are different from the band structure calculations and previous ARPES measurement results.  We observed nearly isotropic superconducting gap around the Fermi surface sheets near $\Gamma$ and M without gap nodes. These rich information will shed more light on the  nature of superconductivity in  the  Fe-based superconductors.

XJZ and MHF thank the funding support from NSFC (Grant No. 10734120 and 10974175) and the MOST of China (973 program No: 2011CB921703 and 2011CBA00103).

$^{*}$Corresponding author: XJZhou@aphy.iphy.ac.cn


\end{document}